\documentclass[longbibliography,aps,prl,twocolumn,nobibnotes,superscriptaddress,floatfix,tightenlines,showpacs,notitlepage]{revtex4-1}
\usepackage{graphicx}
\usepackage{bm}
\usepackage[normalem]{ulem}
\usepackage{amsfonts}
\usepackage{color}
\usepackage{ulem}
\usepackage{amsmath}    
\usepackage{epsfig}
\usepackage{subfigure}  
\usepackage[breaklinks,colorlinks = true,linkcolor = blue,urlcolor=blue,citecolor=blue]{hyperref}
\usepackage{amssymb}
\usepackage{lineno}
\usepackage{hyperref} 
\usepackage{cancel}
\usepackage{comment}

\newcommand{\aver}[1]{ \! \left\langle {#1} \right \rangle \!}

\newcommand{\inv}{\gamma}

\def \rp   {\rm p}
\def \rd   {{\rm d}}

\def \rf   {\rm f}
\def \bu   {\bm {u}}
\def \br   {\bm {r}}

\def \dbu  {\delta \bu}
\def \rh   {\hat{\br}}

\def \taup  {\tau_{\rm p}}
\def \tauL  {\tau_{\rm L}}

\def \mup  {\mu_{\rm p}}

\def \bx {\bm {x}}

\def \diss {\varepsilon}
\def \dissf {\varepsilon_{\rm f}}
\def \dissp {\varepsilon_{\rm p}}

\def \disst {\varepsilon_{\rm t}}

\def \urms  {u_{\rm rms}}
\def \Rey  {\mbox{Re}}
\def \Reeta  {\Rey_\eta}
\def \ueta  {u_\eta}
\def \Deb  {\mbox{De}}
\def \Rel  {\mbox{Re}_{\lambda}}
\def \dur {\delta_r u}

\def \dui {\delta  u_i}

\def \calo {\mathcal{O}}
\def \Stwo {S^2}
\def \Sp 	{S^{\rp}}

\def \Spf 	{\Sp_{\rf}}
\def \Sthree 	{S^3}

\def \Stf 	{S^3_{\rf}}

\def \aij   {\partial_i u_j}

\def \pari 	{\partial_i}

\def \bmf	{\bm{f}}
\def \fluxt	{\Phi_{\rm t}}
\def \fluxf	{\Phi_{\rm f}}
\def \fluxp	{\Phi_{\rm p}}

\def \Pif	{\Pi_{\rm f}}
\def \Pip	{\Pi_{\rm p}}

\def \volume {\Omega}
\def \surf	{\partial \Omega}

\def \expsfpt {\zeta^{\rm PT}_{\rp} }
\def \expsfnt {\zeta^{\rm NT}_{\rp} }
\def \devpt {\delta^{\rm PT}}
\def \devnt {\delta^{\rm NT}}

\def \Ck	{C_{\mathcal{K}}}

\newcommand \lrp[1] {\left( #1 \right)}

\newcommand \bra[1] {\left\langle #1 \right\rangle}
\newcommand \ddr[1] {\frac{\rd #1}{\rd r}}

\newcommand \lrv[1] {\left\lvert #1 \right\rvert}

\begin{document}

\title{The invariant rate of energy extraction by polymers in turbulence}

\author{Alessandro Chiarini}
\email[]{alessandro.chiarini@polimi.it}
\thanks{co-first author}
\affiliation{Complex Fluids and Flows Unit, Okinawa Institute of Science and Technology Graduate University, 1919-1 Tancha, Onna-son, Okinawa 904-0495, Japan.}
\affiliation{Dipartimento di Scienze e Tecnologies Aerospaziali, Politecnico di Milano, via La Masa 34, 20156 Milano, Italy.}
\author{Rahul K. Singh}
\email[]{rksphys@gmail.com}
\thanks{co-first author}
\affiliation{Complex Fluids and Flows Unit, Okinawa Institute of Science and Technology Graduate University, 1919-1 Tancha, Onna-son, Okinawa 904-0495, Japan.}
\author{Marco E. Rosti}
\email[]{marco.rosti@oist.jp}
\affiliation{Complex Fluids and Flows Unit, Okinawa Institute of Science and Technology Graduate University, 1919-1 Tancha, Onna-son, Okinawa 904-0495, Japan.}

\date{\today}

\begin{abstract}
Polymeric turbulence, flows of fluids with dilute polymer additives at high Reynolds numbers, exhibits striking deviations from the Kolmogorovean behaviour of Newtonian turbulence. Recent experiments as well as simulations have uncovered a robust self-similar energy spectrum scaling as $k^{-2.3}$, in sharp contrast to the $k^{-5/3}$ scaling of Newtonian flows. The origin of this novel scaling, however, has remained unresolved. In this work, we uncover the underlying physical mechanism responsible for this emergent behaviour. Using fundamental governing equations aided by scaling arguments, we show that the fluid energy cascade is depleted by the polymers at a constant rate across a wide range of scales. This constant depletion rate acts as a second invariant, alongside the total energy flux, thereby setting the scaling properties of the spectrum. Our results reveal that polymeric turbulence is governed by two simultaneous invariants, unlike the single-invariant structure of Newtonian turbulence, and suggest new strategies for turbulence control through suitably engineered and targeted polymer design.
\end{abstract}

\maketitle

Invariants are central to the physical description of nature and are directly related to underlying symmetries via a fundamental theorem due to Emmy Noether~\citep{Noether1918}. This principle also extends to turbulence of incompressible flows whose evolution is described by the Navier–Stokes equations (NSE). In fully developed turbulence in three space dimensions (3D), which is attained in limit of fluid viscosity $\nu \to 0$, these equations conserve the mean momentum, the energy, and the helicity, corresponding to statistical symmetries under Galilean transformations, time translations, and parity transformations, respectively~\citep{Frisch96}. Additionally, there has been evidence of conformal invariance---an extension of scale invariance by allowing local rescaling---in two-dimensional (2D) turbulence~\citep{Bernard2006}, as well as in 3D rotating turbulence~\citep{Thalabard2011}. Scale invariance simply implies that the NSE are invariant under the transformations $\br \to \lambda \br, \bu \to \lambda^h \bu, t \to \lambda^{1-h} t$ in the inviscid limit, where $\lambda,h \in \mathbb{R}, \br \in \mathbb{R}^3$. For any finite viscosity $\nu >0$, however, invariance can be restored if $\nu \to \lambda^{h+1} \nu$ which leaves the Reynolds number $\Rey \equiv UL/\nu$ invariant, where $U$ is the characteristic velocity at a large scale $L$. Under this rescaling, the average rate of energy dissipation $\bra{\diss} \equiv \nu \langle \lrp{\aij}^2 \rangle$ transforms as $\bra{\diss} \to \lambda^{3h-1} \bra{\diss}$, which implies scale invariance for $h=1/3$~\citep{Benzi1984}. This choice of rescaling underlies Kolmogorov's 1941 (K41) theory of homogeneous, isotropic, incompressible, Newtonian turbulence (NT). In other words, the K41 theory assumes that energy is transferred locally in scales at a constant rate that equals $\bra{\diss}$. This assumption led Kolmogorov to the celebrated 4/5-th law for the third-order velocity structure function $\Sthree (r) \equiv  \langle \lrp{\dbu \cdot \rh}^3 \rangle = -\lrp{4/5} \bra{\diss} r$~\citep{K41c}, where the increments of the velocity fluctuations $\dbu = \bu \lrp{\bx + \br} - \bu \lrp{\bx}$ are computed over a separation $\br$ ($\lrv{\br} = r$) that lies in a suitable, intermediate range of scales; here $\hat{\bm{r}}=\bm{r}/r$. This relation follows directly from the balance of energy detailed by the K\'{a}rm\'{a}n--Howarth relation~\cite{KH1938}. A simple dimensional extension of this relation means that any $\rp$-th order structure function $\Sp (r) \equiv \bra{\lrp{\dbu \cdot \rh}^{\rp}} \sim (\langle \diss \rangle r)^{\rp/3}$~\cite{Frisch96}. Consequently, the energy of fluctuations captured by the second-order structure function scales as $\Stwo (r) \sim \lrp{\langle \diss \rangle r}^{2/3}$ and relates to the energy spectrum $E(k) \sim \langle \diss \rangle^{2/3} k^{-5/3}$ via the Wiener-Khinchin theorem, with $k \sim 1/r$. Thus, Eulerian turbulence structure functions within the K41 phenomenology exhibit a self-similarity with a $\rp/3$ power law $\Sp (\lambda r) \sim \lambda^{\rp/3} \Sp(r)$, so that $\dur \equiv \bra{\lrp{\dbu \cdot \rh}}$ scales with $h = 1/3$. It is now well known that this detailed scale invariance is broken due to intermittency in $\diss$ lending turbulence a multiscaling nature~\citep{Water1991, Frisch96,Sreeni1997,Anirban1998}. However, we do not concern ourselves with intermittency effects in this work.

\begin{figure}
\centering
\includegraphics[width=\columnwidth]{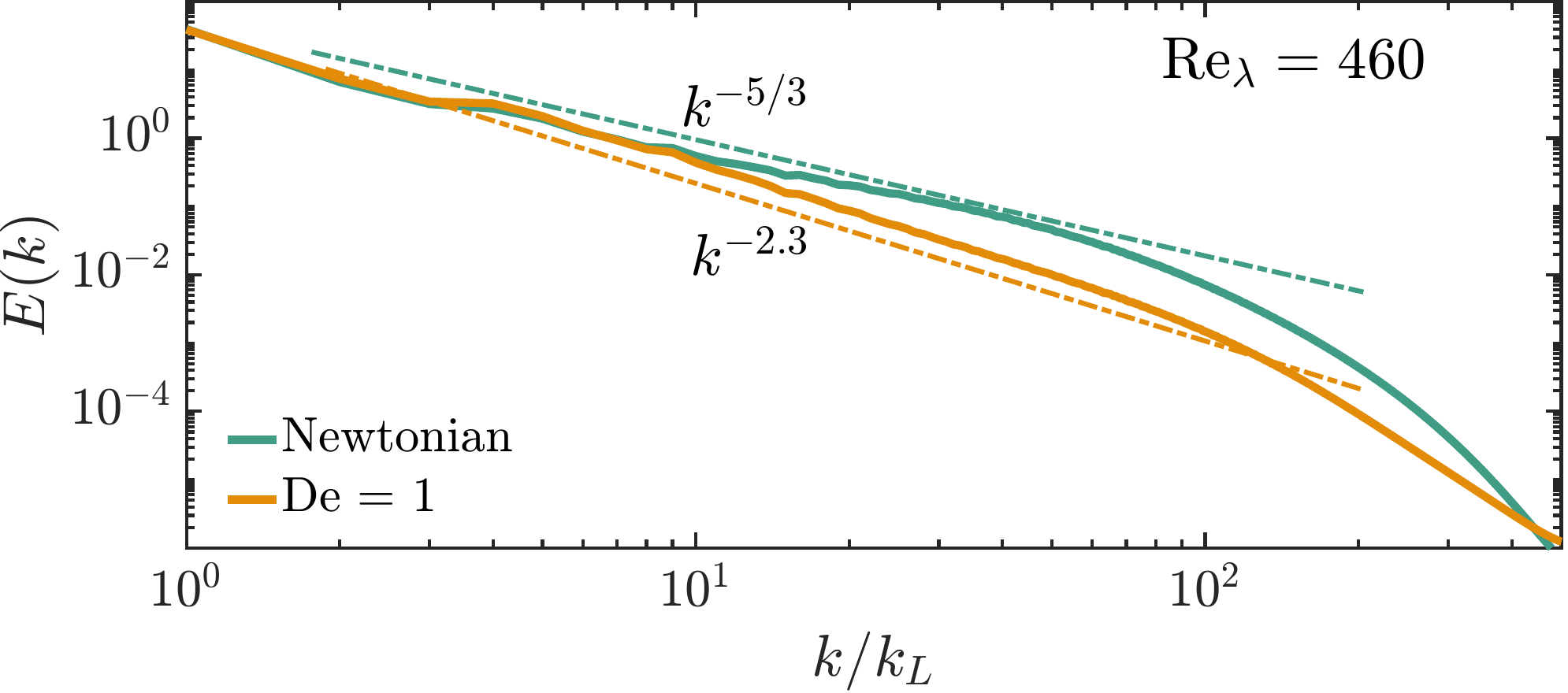}
\caption{Fluid energy spectra for Newtonian and polymeric turbulence at a large $\Rel = 460$ show very distinct self-similar behaviour. The Newtonian $-5/3$rd scaling gives way to a $-2.3$ power-law spectrum in polymeric flows.}
\label{fig:spec1}
\end{figure}

The theory of Kolmogorov, even though only approximately, applies primarily to single-phase Newtonian flows. The introduction of other phases to the flow, such as polymers, significantly alters the characteristics and statistics of turbulence including the distribution of energy across scales~\cite{deangelis-etal-2005,Benzi2018,Bodenschatz21,Marco23}. Polymeric fluid flows, apart from $\Rey$, are additionally characterised by another dimensionless Deborah number $\Deb \equiv \taup/\tauL$, which quantifies polymer elasticity via the relaxation time $\taup$ with respect to a large time scale of the flow $\tauL = L/\urms$, where $\urms$ is root-mean-squared velocity. Recent experimental~\citep{Bodenschatz21} and numerical~\citep{Marco23} studies found that turbulent polymeric flows exhibit a novel self-similarity in an intermediate range of scales $r_a \le r \le r_b$, such that the energy spectrum in polymeric turbulence (PT) scales as $E(k) \sim k^{-2.3}$, a result which has received support from field-theoretic calculations~\citep{Calzetta2024}; see also Fig.~\ref{fig:spec1}. This indicates that the second-order structure function $\Stwo_{\rp}$ in PT exhibits a self-similarity of the form $\Stwo_{\rp} (\lambda r)  \sim \lambda^{1.3} \Stwo_{\rp} (r)$. Although a K41-like description for polymeric turbulence has been proposed~\citep{Ale2024}, the origin of the observed self-similarity and the nature of the associated invariant remain unclear. 

Our aim in this Letter is to uncover the physical phenomenon underlying this emergent scaling. This also means identifying an invariant $\inv$ and a universal constant $\Ck \in \mathbb{R}$ such that:
\begin{align}
	E(k) = \Ck \inv^\alpha k^{-2.3}  ;		\qquad		  k_L \ll k_b \leq k \leq k_a \ll k_\eta ,
\end{align}
where $\alpha \in \mathbb{R} $,  $k_\eta \sim 1/\eta$, $k_L \sim 1/L$, $\eta$ being the smallest scale of the flow.
 This novel self-similarity spans maximal range of scales for $\Deb \sim  \calo(1)$ and $\Rey \to \infty$, while the K41 scaling is recovered in the limits $\Deb \to \lrp{0,\infty}$ and $\Rey \to \infty$~\citep{Marco23}. Hence, we expect that $\inv$ is invariant over a maximal range for $\Deb \sim \calo (1)$ and $\Rey \to \infty$.

We address this problem by considering the simplest case of a homogeneous, isotropic, turbulent polymeric flow in a triperiodic box with side $L = 2\pi$. The flow dynamics is governed by the modified incompressible  NSE ($\partial u_i /\partial x_i = 0$), with an extra stress term that models the interaction of polymers with the carrier fluid, coupled to a simultaneous evolution of the polymer stresses:
\begin{align}
& \frac{\partial u_i}{\partial t} + u_j \frac{\partial u_i}{\partial x_j} = - \frac{1}{\rho} \frac{\partial p}{\partial x_i} + \nu \frac{\partial^2 u_i}{\partial x_j^2} + \frac{1}{\rho} \frac{\partial T_{ij}}{\partial x_j} + f_i \ , \label{eq:NS1} \\
& \frac{\partial R_{ij}}{\partial t} + u_k \frac{\partial R_{ij}}{\partial x_k} = R_{kj} \frac{\partial u_i}{\partial x_k} + R_{ik} \frac{\partial u_k}{\partial x_j} - \frac{ \mathcal{P} R_{ij} - \delta_{ij} }{\taup}. \label{eq:NS2} 
\end{align} 
Here, $\rho$ is the fluid density and $f_i$ is the forcing that sustains statistically stationary turbulence by injecting energy at the large scale $L$. The extra stress tensor $T_{ij}$ is related to the polymer conformation $R_{ij}$ as $T_{ij} = \mup(\mathcal{P} R_{ij} - \delta_{ij})/\taup$, where $\taup$ is the polymeric relaxation time, $\mup$ the polymeric viscosity, and $\delta_{ij}$ is the Kronecker delta. The Peterlin function $\mathcal{P}$ specifies the polymeric model: it selects the purely elastic Oldroyd-B  model for $\mathcal{P}=1$, and the FENE-P model when $\mathcal{P} = \ell^2_{\max} - R_{ii}$, where $\ell_{\max}$ is the maximum polymer extensibility.

In PT, energy injected by the forcing $\bmf$ is, on an average, transferred to smaller scales via dual routes of fluid non-linearity and the fluid polymer interactions. This energy is eventually dissipated away at the smallest scales by both the fluid $\dissf$ and polymeric $\dissp$ phases, such that the total rate of energy dissipation is $\disst = \dissf + \dissp$. In an intermediate range of scales, however, neither forcing nor dissipative effects are important, and the total energy transfer rate $\rd \fluxt/\rd r$ is scale-invariant and equals the total dissipation rate as $\rd \fluxt/\rd r = -(4/3) \bra{\disst}$~\citep{Ale2024}. However, the fluid $\rd \fluxf/\rd r$ and polymeric $\rd \fluxp/\rd r$ contributions to the total transfer rate $\rd \fluxt/\rd r = \rd \fluxf/\rd r + \rd \fluxp/ \rd r$ are themselves non-constant functions of the scale $r$, as can be expected for a non-K41 like behaviour. This can be shown by writing the K\'{a}rm\'{a}n--Howarth--Monin--Hill (KHMH) equation for the balance of the scale energy of the fluctuations $\bra{ \delta u_i^2 }$ \citep{Hill2002,Marati2004,Cimarelli2013,Yasuda2018,Gatti2020,gattere-etal-2023,yao-etal-2024}. The KHMH for PT under the assumptions of statistical stationarity and homogeneity is obtained as~\citep{deangelis-etal-2005,Ale2024}:
\begin{align}
\frac{\partial}{\partial r_j} \bra{ \delta u_j \delta u_i^2 } +
&\frac{\partial}{\partial r_j} \bra{ - 4 \delta u_i T_{ij}^* }  =   2 \nu \frac{\partial^2 \bra{\delta u_i^2}}{\partial r_j^2} \nonumber \\ 
& + 2 \bra{\delta u_i \delta f_i} 
- 4 \bra{\dissf} - 4 \bra{\dissp},
\label{eq:KHMH1}
\end{align}
where $\dissf = - \nu \lrp{\pari u_j}^2$ is the local, positive-definite pseudo-dissipation, $\bra{\dissp} \equiv \aver{ T_{ij} \partial u_i/\partial x_j} = \aver{T_{jj}}/2 \taup$ is the average, positive-definite rate of energy dissipation due to polymers, and the superscript `$*$' denotes the average of any quantity at two points $\bx$ and $\bx + \br$. Now, integrating this scale-by-scale energy balance over spherical volumes $\Omega$ bounded by the surface $\partial \Omega$ of radius $r$ yields the following balance of energy fluxes:
\begin{align}
 \underbrace{  \frac{1}{A} \int_{\surf} \text{d}A \ n_j \bra{\delta u_j \delta u_i^2} }_{\fluxf (r)} &+
\underbrace{ \frac{1}{A} \int_{\surf} \text{d}A \ n_j \bra{ - 4 \delta u_i T_{ij}^*} }_{\fluxp (r)}    \nonumber\\
=  \underbrace{ \frac{2 \nu}{A} \int_{\surf} \text{d}A \ n_j \frac{\partial \aver{ \delta u_i^2}}{\partial r_j} }_{D(r)}		
& +\underbrace{\frac{1}{A} \int_{\volume} \text{d}V \ \bra{ 2 \dui \delta f_i} }_{F (r)} \nonumber \\
&- \frac{4}{3} \underbrace{ \left( \bra{ \dissf } + \bra{ \dissp } \right) }_{\bra{\disst}} r,
\label{eq:KHMH}
\end{align}
where $A = 4 \pi r^2$ is the area of the surface described by $\surf$, and $\bm{n}$ is the outward unit vector normal to the spherical surface. Note that, the nonlinear flux $\fluxf$ and the third-order fluid structure function $\Stf$ are related as $\Stf(r) = (3/5) \fluxf (r)$, under the assumption of homogeneity, isotropy, and incompressibility; see the Supplemental Material (SM). Now, in the intermediate range of scales, i.e., in the limits of $\Rey \to \infty$ and $r/L \to 0$ in that order, Eqn.~\eqref{eq:KHMH} can be reduced further to
\begin{align}
\fluxt \equiv \fluxf (r) + \fluxp (r) = - \frac{4}{3} \bra{\disst} r,
\label{eq:KHMH_in}
\end{align}
where, as anticipated, the total flux $\fluxt$ has a simple, exact scaling form similar to $\Stf$ (or $\fluxf$) in the 4/5th Kolmogorov law for NT. The distinction between NT and PT is also clearly brought out: the flux of total energy in PT is a cumulative contribution of $\fluxf$ and $\fluxp$, the latter of which is absent in NT. Similarly, the dissipation of energy in PT is also a cumulative of $\dissf$ and $\dissp$, compared to only $\dissf$ in NT. This is more evident from the scale-by-scale balance of total energy transfer, obtained by differentiating Eqn.~\eqref{eq:KHMH_in} with respect to $r$:
%
%
\begin{align}
	\Pif (r) + \Pip (r) \equiv \ddr{\fluxf} + \ddr{\fluxp} = - \frac{4}{3} \bra{\disst}.
	\label{eq:der1}
\end{align}
\begin{figure}
	\centering
	\includegraphics[width=0.49\textwidth]{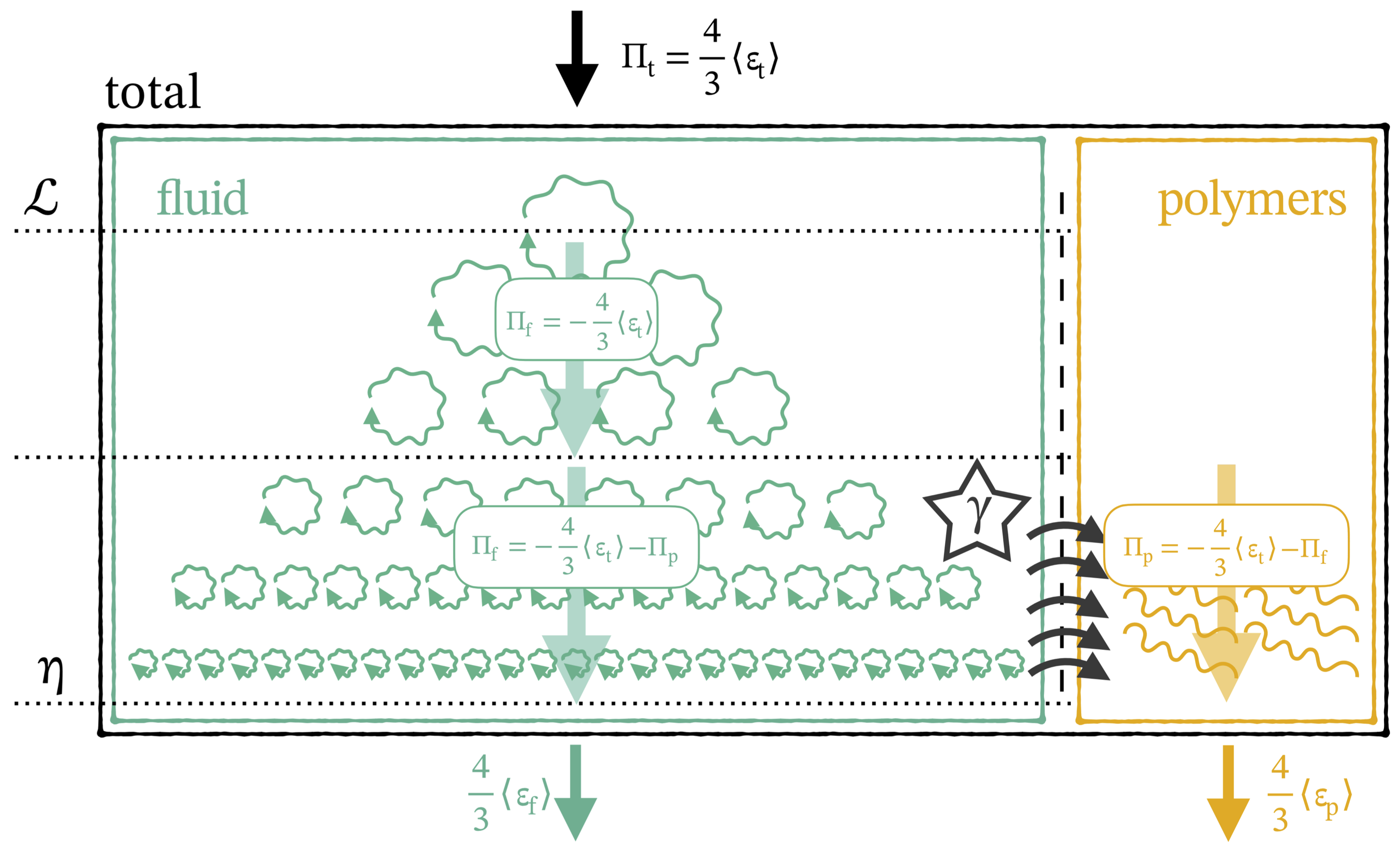}
	\caption{Sketch of the modified energy cascade in the presence of the polymeric additives. In the range of scales where polymers are active, two different transfer processes coexist: one described by the classical Richardson scenario while the other is associated with the polymeric microstructure. In this range, a constant flux is lost from the classical cascade route, and transferred to small scales by the polymeric microstructure.}
	\label{fig:sketch}
\end{figure}
In NT, $\disst = \dissf$ and $\fluxp = 0$ so that $\Pif(r) = - (4/3) \bra{\dissf}$ is an invariant across an intermediate range of scales~\citep{Frisch96,pope-2000}. Eqn.~\eqref{eq:der1} shows that in PT there is a finite, non-Newtonian contribution $\Pip (r) \neq 0$ at all scales. Moroever, the possibility of a constant $\Pip$ is ruled out as that would again imply a K41-like scaling, which contradicts the observation in Fig.~\ref{fig:spec1}. Hence, the emergence of the new scaling in PT requires that $\Pif(r), \Pip(r) \neq const.$ within the intermediate range of scales with the constraint
\begin{align}
\ddr{\Pif} = - \ddr{\Pip},
\label{eq:der2}
\end{align}
following from Eqn.~\eqref{eq:der1}. This equality simply shows that, at any scale in the intermediate range, there is a net loss of flux from the fluid mode to the polymers, which we sketch in Fig.~\ref{fig:sketch}. However, Eqn.~\eqref{eq:der2} does not yet prescribe the exact nature of this flux loss in the form of an $r$-dependence. This cannot be derived from the governing equations themselves, and can only be provided based on physical arguments and/or data from numerical/experimental observations.

As a minimal model, one may assume that polymers extract energy from the fluid cascade at a constant rate across the intermediate scales, i.e., $\Pif' = -\Pip' = - \inv = const $, where primes denote derivatives with respect to $r$. While we cannot \textit{a priori} exclude alternative prescriptions in which $\Pif'$ and $\Pip'$ are non-trivial functions of $r$, the assumption of a constant is supported by both experiments~\citep{Bodenschatz21} and numerical simulations~\citep{Marco23}, which report a second-order structure function $S^2_{\rf} \sim r^{1.3}$ in PT. The use of a simple dimensional argument then yields $\Stf \sim \lrp{S^2_{\rf}}^{3/2} \sim\lrp{ r^{1.3}}^{3/2} = r^{1.95} \approx r^2$. Thus, $\Pif' = \fluxf'' \sim \lrp{\Stf}''= const$. Therefore, we can modify Eqn.~\eqref{eq:der2} to now identify the scale-by-scale rate of flux transfer as the invariant $\inv$ in the limit $\Rey \to \infty$, $\Deb \sim \calo (1)$, $r/L \to 0$ and for $r_a \le r \le r_b$:
\begin{align}
 \ddr{\Pif} = - \ddr{\Pip} = -\inv.
	\label{eq:inv}
\end{align}
This relation is the fundamental, defining relation for PT, much like Kolmogorov's $4/5$th law in NT, $\rd \Stf/ \rd r = - (4/5) \langle \dissf \rangle$. The validity of this relation depends on the proximity to these limiting regimes. For instance, the range of scales over which $\inv$ is invariant shrinks if either $\Rey$ is small or $\Deb \neq 1$~\citep{singh-rosti-2024}. More precisely, $\Pif' (r) = -\inv $ for $ \eta \ll r_a \le r \le r_b \ll L$ with $r_{a,b} = r_{a,b} (\Rey,\Deb)$. It is also worth stating that $\inv$ is an exchange invariant that emerges due to the presence of multiple phases in the flow, and is \textit{not} an inviscid invariant of PT.

Of course, Eqn.~\eqref{eq:inv} implies the relation $\Stf \sim -\inv r^2$, 
%
%
%
which can be extended to any $\rp$-th order structure function as $S^{\rp}_{\rf}(r) \sim \inv^{\rp/3} r^{2\rp/3}$, disregarding intermittency corrections. In particular, this means $S^2_{\rf}(r) \sim \inv^{2/3} r^{4/3}$ and correspondingly $E(k) \sim \inv^{2/3} k^{-7/3}$, which is in agreement with the simulation data shown in Fig.~\ref{fig:spec1}. More precisely, in the limit of $\Rey \to \infty$, $\Deb = 1$, $r/L \to 0$ (taken in that order), the velocity fluctuation statistics in PT admit a universal, self-similar scaling form determined by an invariant rate of flux transfer $\inv$ as
\begin{subequations}
  \begin{align}
&    \Spf \lrp{r} = C_{\rp} \inv^{\rp/3} r^{2\rp/3} \ ;	\quad \eta \ll r_a \le r \le r_b \ll L,	\label{eq:statsr} \\
 &  E(k) = \Ck \inv^{2/3} k^{-7/3}  \ ;	  \ \ k_\eta \gg k_a \ge k \ge k_b \gg k_L, \label{eq:statsf} 
  \end{align}
\label{eq:ptstats}
\end{subequations}
for some universal, dimensionless constants $C_{\rp}, \Ck \sim \calo (1)$, which are the analogous of the well known Kolmogorov constants in NT~\cite{Sreeni1995,Yeung1997,Donzis2010}. We discuss some immediate implications of identifying the novel invariant $\inv$ and its relation with $\bra{\dissp}$ alongside describing model PT in detail in the End matter.


%
\begin{figure*}[!ht]
	\centering
	\includegraphics[width=\textwidth]{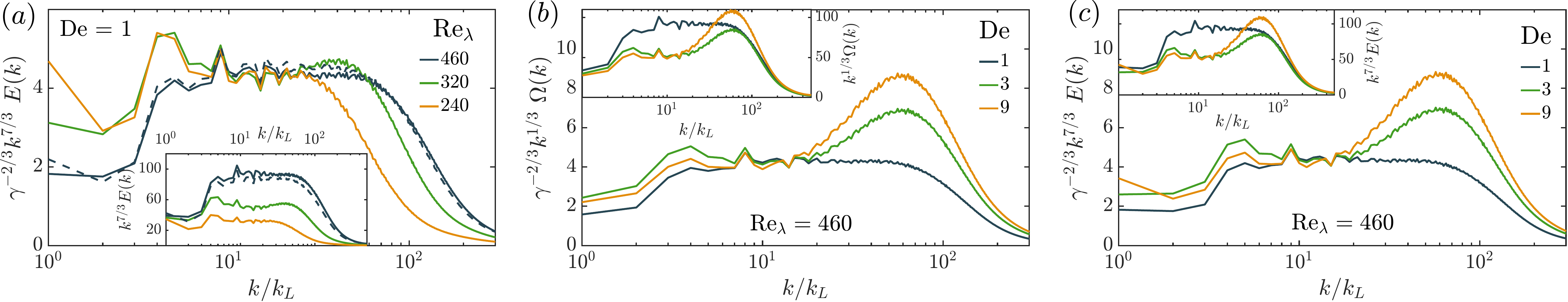}
	\caption{(a) Compensated energy spectra $\inv^{-2/3} k^{7/3}  E(k)$ in the main panel for different $\Rel \in \{240, 320, 460 \}$ but the same $\Deb = 1$ show a remarkale collapse. The dashed spectrum is from a FENE-P simulation, while the solide curves are for Oldroyd-B. A compsensation by only wavenumbers in the inset is devoid of any collapse. (b) The compensated enstrophy spcetrum $\inv^{-2/3} k^{1/3}\Omega(k)$ for $\Rel = 460$ and $\Deb \in \{ 1, 3, 9 \}$. Again, merely a compensation with scales does not leaad to a collapse in the inset. (c) Compensated energy spectra for same parameters as (b). Clearly, the PT regime shrinks as $\Rel$ decreases and $\Deb$ deviates from 1.}
	\label{fig:spec}
\end{figure*}

This concludes our theoretical model for PT, which we now test against data from direct numerical simulations (DNS). To this end, we analyse Fourier-space statistics, which provide a broader and clearer scaling regime. The data are obtained from DNS of PT in an Oldroyd-B fluid, with the microscale Reynolds number defined as $\Rel = u_{\text{rms}} \lambda/\nu$, where $\lambda$ is the Taylor microscale. We explore the ranges $\Rel \in \{ 240,320,460 \}$ and $\Deb \in \{1,3,9\} $. To verify model-independence, we also perform simulations using the FENE-P model at $\Rel = 460$ and $\Deb=1$. Details of the numerical methods and simulation setup are provided in the SM. 

The proposal of a new invariant $\inv$ is at the core of our model description of PT, which guarantees the relation~\eqref{eq:statsf} for the energy spectrum. This can be rewritten in a compensated form as
\begin{align}
		 \inv^{-2/3} k^{7/3} E(k)  =  \Ck     \ ;&	\qquad 	k_a \ge k \ge  k_b .	\label{eq:compstatsf}
\end{align}
This implies that the appropriately compensated energy spectrum should exhibit a scale-independent plateau---i.e. a universal constant---within the range of validity of the theory. As discussed earlier, this range narrows with decreasing $\Rel$ and depends non-monotonically on $\Deb$, being widest at $\Deb \sim \mathcal{O}(1)$~\citep{Marco23}. Consequently, the compensated spectra for different $\Rel$ and $\Deb$ should collapse onto the same constant value $\Ck$ over an intermediate range that contracts with decreasing $\Rel$ and for $\Deb \nsim \mathcal{O}(1)$.

We begin the model validation with a few consistency checks. First, we verify that Eqn.~\eqref{eq:statsf} correctly predicts the polymeric energy spectrum scaling $E(k) \sim k^{-7/3}$ for PT, as shown in Fig.~\ref{fig:spec1} for $\Deb =1$ and $\Rel = 460$. (alongwith the Newtonian spectrum with $E(k) \sim k^{-5/3}$.) Therefore, the scale-compensated energy spectrum $k^{7/3} E(k)$ should exhibit a flat region over an intermediate range of scales. This is evident in the insets of Fig.~\ref{fig:spec} whose panels $(a), (c)$ respectively show the energy spectra for $\Rel = \{240,320,460\}$ at $\Deb=1$ and $\Deb \in \{1,3,9\}$ at fixed $\Rel = 460$. Panel (b) plots the enstrophy spectra $\Omega (k)$ for $\Deb \in \{1,3,9\}$ at fixed $\Rel = 460$ whose scaling behaviour can be estimated using the following simple arguments. 
 We know that velocity fluctuations up to a wavenumber $k$ scale as $u^2_k \sim k E(k) \sim \inv^{2/3} k^{-4/3}$, implying $u_k \sim \inv^{1/3} k^{-2/3}$. Therefore the vorticity fluctuations scale as $\omega_k \sim k u_k \sim \lrp{\inv k}^{1/3}$. Enstrophy spectrum can then be estimated as $k \Omega (k) \sim \omega_k^2 \sim \lrp{\inv k}^{2/3}$ so that $\Omega (k) \sim \inv^{2/3} k^{-1/3}$. The inset of panel $(b)$ clearly shows a flat intermediate range upon a scale compensation $k^{1/3} \Omega (k)$. These plots moreover confirm that the extent of the scaling range diminishes with decreasing $\Rel$ as well as when $\Deb$ deviates further from from unity. Model independence is demonstrated by the FENE-P spectrum (dashed blue curve) which closely matches the Oldroyd-B result (solid blue curve)~\citep{Marco23,singh-rosti-2024}. Thus, we confirm that PT exhibits the predicted $E(k) \sim k^{-7/3}$ scaling, whose extent is a function of $(\Rel, \Deb)$ and is maximal in the limit $\Rel \rightarrow \infty$ and $\Deb = 1$.

We now turn to the core validation of our conjectures. Specifically, we demonstrate that a further compensation by $\inv$, as suggested by Eqn.~\eqref{eq:compstatsf}, leads to a collapse of the spectra across different $\Rel$ and $\Deb$. To this end, we estimate $\inv$ using Eqn.~\eqref{eq:inv} for each $(\Rel,\Deb)$ pair by computing the flux derivatives $\Pif'(r)$ and $\Pip'(r)$. Additional computational details and values of $\inv$ are provided in the SM. We also show that independent estimates of $\inv$ using Eqn.~\eqref{eq:inv2} and relating it to $\bra{\dissp}$ yield consistent values.
Using these estimates, we observe that the fully compensated spectra $\inv^{-2/3} k^{7/3} E(k) $ and $\inv^{-2/3} k^{1/3} \Omega(k)$ collapse remarkably well across different $(\Rel,\Deb)$ values, as shown in the main panels of Fig.~\ref{fig:spec}. The intermediate plateau for $E(k)$ in panels (a) and (c) yields a universal constant $\Ck = 4.1 \pm 0.4$, consistent with our prediction that $\Ck = \mathcal{O}(1)$. 
This confirms our conjecture~\eqref{eq:ptstats} that in the range $ \eta \ll r_a \le r \le r_b \ll L$ the fluid statistics are governed by the novel invariant $\inv$, provided that $\Rel \rightarrow \infty$ and $\Deb \rightarrow 1$. The collapse of $\inv^{-2/3} k^{1/3} \Omega (k)$ for $\Rel =460$ and $\Deb = 1,3,9$ using the same values of the invariant $\inv$ provides additional confirmation of our results. We further discuss some immediate implications of $\inv$ in section 6 of SM.

In conclusion, our conjectures, supported by data and heuristic analysis, significantly advance the current understanding of PT by uncovering a novel invariant, denoted $\inv$. This invariant underlies a new form of self-similarity---distinct from that observed in Newtonian turbulence---and captures a simple yet fundamental mechanism: a constant rate of flux transfer from the carrier fluid to  polymers. Together with the total dissipation rate $\bra{\disst}$~\citep{Ale2024}, the invariant $\inv$ provides a complete statistical description of PT, suggesting that any theoretical framework must incorporate both quantities. 
This highlights a fundamental distinction from NT where the statistics are governed by only a single invariant $\bra{\dissf}$~\citep{K41c,Frisch96}.
Nonetheless, a rigorous derivation from first principles along with validation through closure models and numerical or experimental evidence across diverse flow configurations is essential to firmly establish the universality of the constant $C_\mathcal{K}$ and further substantiate our results.

Our work lays new ground toward a more fundamental understanding of PT in particular, and turbulence in complex fluids in general. The discovery of a novel invariant governing the rate of energy flux transfer has practical implications for controlling turbulence via polymers of tailored elasticity. In particular, when the polymer relaxation time matches the largest turbulent timescales, energy can be extracted in a controlled manner across a broad range of scales, enabling a tunable mechanism for modulating turbulence.
The discovery of this invariant also points to enhanced separation of Lagrangian tracers in PT, as captured by the second-order Lagrangian structure function. Specifically, we expect $S_2 (\tau) \equiv \bra{|\bu (t+\tau) -\bu(t)|^2 }  \sim \inv^2 \tau^4$ in comparison to NT where $S_2 (\tau) \sim \langle \dissf \rangle \tau$~\cite{SreeniRev2025}. This suggests significantly stronger mixing in PT with important implications for transport processes in viscoelastic flows, an aspect that warrants further investigation. 
%
%

Finally, we remark that the spectra in Fig.~\ref{fig:spec} exhibit a distinct bump close to the onset of the dissipation range. This phenomenon---known as the bottleneck---has been extensively studied in NT~\citep{Falkovich1994,Lohse1995,Yeung1997,Brandenburg2003,Meneveau2008,SSR2008,SSR2013}, and is attributed to incomplete thermalisation at large wavenumbers~\cite{SSR2008}. While a detailed analysis lies beyond the scope of this work, we note that the bottleneck is entirely suppressed at $\Deb = 1$ while re-emerging as $\Deb$ deviates from unity. This suggests that polymer elasticity can also modulate the thermalisation in the Navier--Stokes dynamics.
Thus, a key question in the non-equilibrium statistical mechanics of turbulence is how do polymers inhibit thermalisation at small scales? While this has been explored in simpler systems like Burgers’ equation~\cite{SSR2011,Sugan2020}, more remains to be understood with regards to suppressing thermalisation in Navier--Stokes turbulence. We believe our findings offer a stepping stone toward resolving this open problem, and lay the groundwork for a deeper understanding of polymeric turbulence and, more broadly, the behaviour of far from equilibrium complex fluids.

\begin{acknowledgments}
\section*{Acknowledgments}
The research was supported by the Okinawa Institute of Science and Technology Graduate University (OIST) with subsidy funding to M.E.R. from the Cabinet Office, Government of Japan. M.E.R. was also supported by funding from the Japan Society for the Promotion of Science (JSPS), grant 24K17210 and 24K00810. The authors acknowledge the computer time provided by the Scientific Computing section of Research Support Division at OIST and by HPCI, under the Research Project grants \textit{hp210269}, \textit{hp220099}, \textit{hp230018}, and \textit{hp250035}.
\end{acknowledgments}

\bibliography{refs,../Wallturb}

\section{End Matter}

\textbf{\textit{Appendix A: Alternate Interpretation of $\inv$}} --- 
An alternate way to interpret the novel invariant $\inv$ is by relating it to the polymer dissipation rate $\bra{\dissp}$. Our discussion in the main text interprets $\inv$ as the scale-by-scale invariant rate of flux transfer from the fluid to the polymers over an intermediate range of scales. Now, in a statistically stationary state of PT, the resulting net transferred energy must be completely dissipated away by the polymers per unit time. This interpretation also follows directly from the definition of $\inv$ in Eq.~\eqref{eq:inv}, as:
\begin{align}
	\ddr{\Pip} = \inv  \implies 	 \Pip (r_b) - \Pip(r_a) = \inv \Delta \ell,			\label{eq:inv1}
\end{align}
where $\Delta \ell = r_b - r_a$ denotes the bandwidth over which $\inv$ remains constant.
Assuming elastic effects dominate the scales $0 \leqslant r \leqslant r_b$ (i.e., $r_a \rightarrow 0$), while purely Newtonian dynamics prevail for $r_b \leqslant r \leqslant L$, we consider the idealised scenario depicted in Fig.~\ref{fig:sketch}. Accordingly, we set $\Pip(r) = 0 $ for $r_b \leqslant r \leqslant L$, while at the smallest scales we have that $\lim_{r \rightarrow 0} \Pip(r) = - (4/3) \bra{\dissp}$, since energy can only be dissipated by polymers and cannot be transferred to yet smaller scales. 
Applying these boundary conditions to Eq.~\eqref{eq:inv1} gives the expected relation:
\begin{align}
	\inv \Delta \ell = \frac{4}{3} \bra{\dissp}	\Rightarrow  \inv =  \frac{4}{3} \frac{\bra{\dissp}}{\Delta \ell} \approx \frac{4}{3} \frac{\bra{\dissp}}{r_b} \ ; \quad r_b \gg r_a .
	\label{eq:inv2}
\end{align}
Thus, the new invariant $\inv$ can also be estimated as the ratio of polymer dissipation to the bandwidth of the novel scaling regime in PT. (We provide eviednce from data in the SM.)

\textbf{\textit{Appendix B: Implications of $\inv$}} --- A direct consequence of identifying $\inv$ as the invariant parameter is the steeper growth (decay) of velocity fluctuations in PT as compared to NT. Specifically, the structure function scaling exponents $\zeta_{\text{p}}$ in PT are predicted to be twice those in NT, i.e., $ 2\rp/3$ versus $\rp/3$. This also implies that intermittency corrections are given as $\devpt = \expsfpt - 2\rp/3$ in PT, while we have $\devnt = \expsfnt - \rp/3$ in NT. We note that previous numerical studies have found these intermittency corrections to be approximately equal~\citep{Marco23}.
Now, since $\inv$ is the rate of energy loss from the fluid mode per unit time at each scale, it can be estimated as $\inv \sim u_r^2 /(r \tau_{f,r}) \sim u_r^3/r^2$, where $\tau_{f,r} \sim r/u_r$ is the typical lifetime of velocity fluctuations $u_r$ at scale $r$. Thus, under the transformations $\br \to \lambda \br, \bu \to \lambda^h \bu$, and $t \to \lambda^{1-h} t$, we have that $\inv  \to \lambda^{3h-2} \inv$ remains scale invariant when $h = 2/3$. 
The equivalent `Kolmogorov scale' quantities in PT can be determined using the same principles as one follows for Newtonian turbulence. The Kolmogorov scale in NT is defined as the scale at which the $\Rey$ number is one, i.e. $\Rey_\eta =1$. Using the same definition to obtain the equivalent `Kolmogorov scale' for PT yields:
\begin{align}
	\Reeta = \frac{\ueta \eta}{\nu} = 1		\implies    \frac{\inv^{1/3} \eta^{2/3} \eta }{\nu} = 1  	\implies 	\eta = \lrp{\frac{\nu^3}{\inv}}^{1/5}
\end{align}
This means that the corresponding velocity and time scales are given as:
\begin{align}
	\ueta = \inv^{1/3} \eta^{2/3} = \lrp{\nu^2 \inv}^{1/5} \ ; \qquad \tau_\eta = \frac{\eta}{\ueta} = \lrp{\frac{\nu}{\inv^2}}^{1/5}
\end{align}
A further consequence is the dependence of required grid resolution $N^3$ on the large scale $\Rey$ for a purely PT flow. This can be obtained by considering the large scale velocity $U$ and length scale $L$:
\begin{align}
	&	\frac{\Rey}{\Reeta} = \frac{UL/\nu}{\ueta \eta/\nu} = \frac{U}{\ueta} \frac{L}{\eta} = \lrp{\frac{L}{\eta}}^{5/3}		\nonumber \\
	&	N^3 \sim \lrp{\frac{L}{\eta}}^{3} \sim   \Rey^{9/5}.
\end{align}
This is smaller than the expected growth of resolution in NT where $N^3 \sim \Rey^{9/4}$\citep{Frisch96}.

\textbf{\textit{Appendix C: Model Description}} --- A straightforward interpretation of the multiscaling behaviour observed in polymeric turbulence is illustrated schematically in Fig.~\ref{fig:sketch}. Recent studies \citep{Marco23} have indeed shown that PT exhibits Kolmogorov-like scaling at large scales ($r_b \le r \ll L$), transitioning to the novel elastic-dominated regime at smaller scales ($\eta \ll r_a \le r \le r_b$).
A forcing $F$ acting at the largest scale $L$ maintains a stationary state of PT by injecting energy into the flow at an average rate given by $(3/4) \rd F/\rd r = \bra{\disst} \equiv \bra{\dissf} + \bra{\dissp}$. This energy is then transferred to smaller scales by the fluid nonlinear cascade as well as fluid-polymer interactions. 
The larger scales, $r_b \le r \ll L$, remain largely unaffected by polymers and are dominated by the fluid cascade, since the local fluctuation time scale $\tau_{f,r}$ exceeds the polymer relaxation time, i.e. $\tau_\text{p}/\tau_{f,r}<1$.
In this range, $\Pip(r) \approx 0$ giving $\Pif(r) \approx - (4/3)\bra{\disst}$, which means the Newtonian, K41 results hold: $\Stwo_{\rf} (r) \sim \bra{\disst}^{2/3} r^{2/3}$ and $E(k) \sim \bra{\disst}^{2/3} k^{-5/3}$.
At smaller scales, $\eta \ll r_a \le r \le r_b$, fluid-polymer interactions become significant, and the nonlinear cascade is depleted at a constant rate $\inv$ in favour of the polymer-mediated energy transfer. In this regime, the polymer relaxation time exceeds the local fluctuation time scale, i.e., $\tau_\text{p}/\tau_{f,r} \ge 1$.
Here, Eqn.~\eqref{eq:inv} holds, yielding the new $\Stwo_{\rf} (r) \sim \gamma^{2/3} r^{4/3}$  and $E(k) \sim \gamma^{2/3} k^{-7/3}$ scalings. 
In the scenario where polymers influence the entire range of scales, i.e. $r_b \approx L$, the classical K41 scaling no longer holds at any scale, and the novel polymer-driven scaling extends across the entire inertial range $\eta \ll r \ll L$. This behaviour is clearly observed in our simulations, as shown in Fig.~\ref{fig:spec1}. 
At the smallest scales, energy is ultimately removed from the system via viscous and polymeric dissipation, quantified by $\bra{\dissf}$ and $\bra{\dissp}$, respectively.

\newpage
\section*{Supplemental Material}

\subsection{The governing equations and the numerical database}

This section provides information on the physical model considered in the present study. We first introduce the governing equations for the polymeric flow, and then we provide an overview of how these equations are discretised and advanced in time.

We consider an incompressible turbulent flow with polymeric additives. The governing equations for a dilute polymeric solution are the incompressible Navier--Stokes equations, with an additional polymeric stress term that models the back reaction of the polymers on the flow, coupled with an evolution equation for the polymer stresses, i.e.,
\begin{equation}
\begin{gathered}
\frac{\partial u_i}{\partial t} + u_j \frac{\partial u_i}{\partial x_j} = - \frac{1}{\rho} \frac{\partial p}{\partial x_i}
+ \nu \frac{\partial^2 u_i}{\partial x_j \partial x_j} + \frac{1}{\rho} \frac{\partial T_{ij}}{\partial x_j} + f_i, \\
\frac{\partial R_{ij}}{\partial t} + u_k \frac{\partial R_{ij}}{\partial x_k} = R_{kj} \frac{\partial u_i}{\partial x_k} + R_{ik} \frac{\partial u_k}{\partial x_j} - \frac{\left( \mathcal{P} R_{ij} - \delta_{ij} \right)}{\tau_\text{p}}, \\
\frac{\partial u_j}{\partial x_j} = 0;
\end{gathered}
\label{eq:NS}
\end{equation}
here $\rho$ is the fluid density, $p$ is the pressure, $\nu$ is the fluid kinematic viscosity, and $f_i$ is the external forcing used to sustain the flow. In the momentum equation the presence of the polymers is accounted for by means of the extra-stress tensor $T_{ij}$, which is related to the conformation tensor $R_{ij}$ as
\begin{equation}
T_{ij} = \mu_\text{p} \frac{ \mathcal{P} R_{ij} - \delta_{ij} }{\tau_\text{p}},
\end{equation}
where $\tau_\text{p}$ is the polymeric relaxation time, $\mu_{\text{p}}$ the polymeric viscosity, and $\delta_{ij}$ is the Kronecker delta. $\mathcal{P}$ is the Peterlin function, which is equal to $\mathcal{P}=1$ for the purely elastic Oldroyd-B model and to $\mathcal{P} = (\ell_{max}^2 -3)/(\ell_{max}^2-R_{ii})$ for the FENE-P model, where the polymers are modelled as Finitely Extensive Nonlinear Elastic springs with the Peterlin approximation; $\ell_{max}$ is the maximum polymer extensibility, and $R_{ii}$ represents the instantaneous measure of the end-to-end length of the polymers.

In this work, we use the DNS database that has been introduced by \citet{singh-rosti-2024}. Here we only briefly recall the numerical method and the related computational procedures, and we refer to their paper for full details. The governing equations are integrated in time using the in-house solver Fujin, which is based on an incremental pressure-correction scheme. The equations are solved on a staggered grid using second-order finite-differences in all the three directions. We use a second-order Adams-Bashforth time scheme for the velocity field and a second-order Crank-Nicolson scheme for the non Newtonian stress tensor. A log-conformation formulation \citep{izbassarov-etal-2018} is used to ensure that the conformation tensor $R_{ij}$ is positive-definite at all times. We sustain turbulence using the Arnold-Beltrami-Childress (ABC) cellular-flow forcing \citep{podvigina-pouquet-1994}.

We consider a cubic box of size $L=2\pi$ having periodic boundary conditions in all the directions. The domain is discretised with $N^3=1024^3$ grid points to ensure that (for all cases) all the scales down to the smallest dissipative ones are properly resolved, i.e., such that $\eta/\Delta x \le 1$ where $\eta$ is the Kolmogorov length scale and $\Delta x$ the grid spacing. The parameters of the simulations are chosen to achieve in the purely Newtonian case a Taylor-microscale Reynolds number of $\text{Re}_\lambda = u_{\text{rms}} \lambda/\nu \in \{ 240,320,460\}$, where $u_{\text{rms}}$ is the root mean square of the velocity fluctuations and $\lambda$ the Taylor length scale. The Deborah number $\text{De}= \tau_\text{p}/\tau_\text{f}$, where $\tau_\text{f} = L/u_{\text{rms}}$ is the turnover time of the largest eddies of the flow, is varied in the range $\text{De} \in \{1,3,9\}$. In all the considered cases, the fluid ($\mu_\text{f}$) and polymer ($\mu_\text{p}$) viscosities are fixed such that $\mu_\text{f}/(\mu_\text{f} + \mu_\text{p})=0.9$. The simulations are advanced in time with a constant time step of 
$\Delta t \approx 2 \times 10^{-3} \tau_\eta$, with $\tau_\eta$ being the characteristic Kolmogorov time scale. For all cases we have considered an Oldroyd-B fluid, but for $\text{Re}_\lambda = 460$ and $\text{De}=1$ we have also performed an additional simulation using the FENE-P model, to show the independence of the results on the polymeric model.

\subsection{The relation between $\fluxf$ and $S_3$}
\label{sec:S3Phi}

In this section, we show that Eqn.
\begin{equation}
\frac{\text{d}\Phi_\text{f}}{\text{d} r} = - \frac{4}{3} \langle \varepsilon_\text{f} \rangle r
\label{eq:phif}
\end{equation}
is equivalent to the celebrated $4/5$ Kolmogorov law, which relates the third-order structure function $S_3 \equiv \aver{ \left[ \left( u_i (\bm{x} + \bm{r}) - u_i(\bm{x}) \right) r_i/|\bm{r}| \right]^3}$ with the average fluid dissipation $\aver{\varepsilon_\text{f}}$ in the inertial range of scales. For simplicity, we consider a purely Newtonian flow, i.e., $\varepsilon_\text{p}=0$ and $\Phi_\text{p}=0$. By recalling the definition of $\Phi_\text{f}$ (see Eqn. 5 of the main text) and using the assumption of isotropy, we can write
\begin{equation}
  \Phi_\text{f}(r) = \frac{1}{A} \int_{\partial \Omega(r)} \text{d}A n_j \langle \delta u_j \delta u_i^2 \rangle = \langle \delta u_\parallel \delta u_i^2 \rangle(r),
\end{equation}
where the $\cdot_\parallel$ subscript denotes the direction parallel to $\bm{r}$. As a result, Eqn.~\eqref{eq:phif} can be simply recast as
\begin{equation}
  \langle \delta u_\parallel \delta u_i^2 \rangle = - \frac{4}{3} \langle \varepsilon_\text{f} \rangle r.
\end{equation}
At this point, we exploit the incompressibility constrain which implies \citep{hill-1997,hill-2002} that
\begin{equation}
r \frac{\partial \langle \delta u^3_\parallel \rangle}{\partial r} + \langle \delta u^3_\parallel \rangle - 6 \langle \delta u_\parallel \delta u^2_\perp \rangle = 0,
\end{equation}
(here the $\cdot_\perp$ subscript denotes the direction perpendicular to $\bm{r}$) and write
\begin{equation}
  \frac{r}{3} \frac{\partial \langle \delta u^3_\parallel \rangle}{\partial r} +
  \frac{4}{3} \langle \delta u^3_\parallel \rangle = 
  - \frac{4}{3} \langle \varepsilon_\text{f} \rangle r,
  \label{eq:du3}
\end{equation}
where $ \aver{\delta u^3_\parallel } \equiv  \Stf(r)$. We now multiply Eqn.~\eqref{eq:du3} by $3r^3$ and integrate between $0$ and $r$ to obtain
\begin{equation}
  \langle \delta u^3_\parallel \rangle(r) \equiv \Stf (r) = - \frac{4}{5} \langle \varepsilon_\text{f} \rangle r,
\end{equation}
which is indeed the $4/5$ Kolmogorov law.

\subsection{The scale-by-scale energy budget}

\begin{figure}
\centering
\includegraphics[width=0.49\columnwidth]{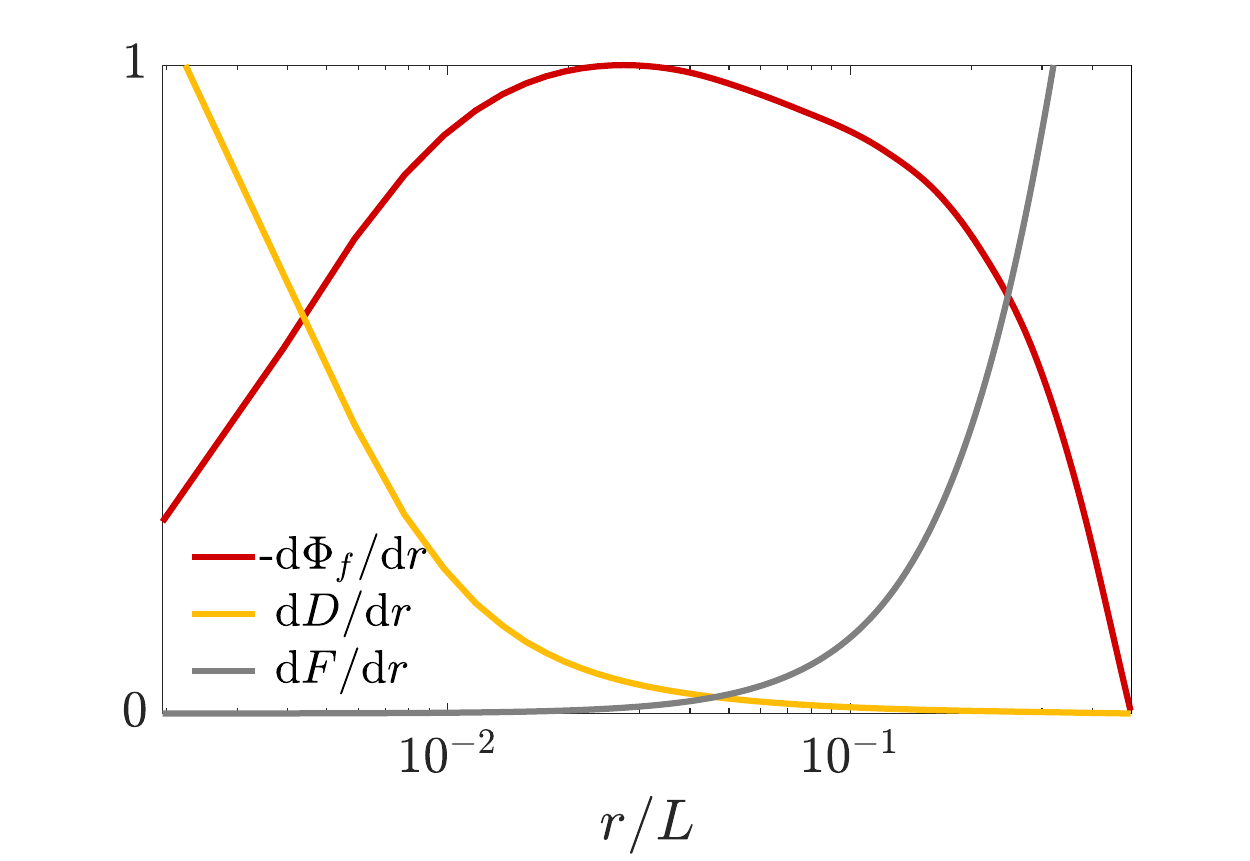}
\includegraphics[width=0.49\columnwidth]{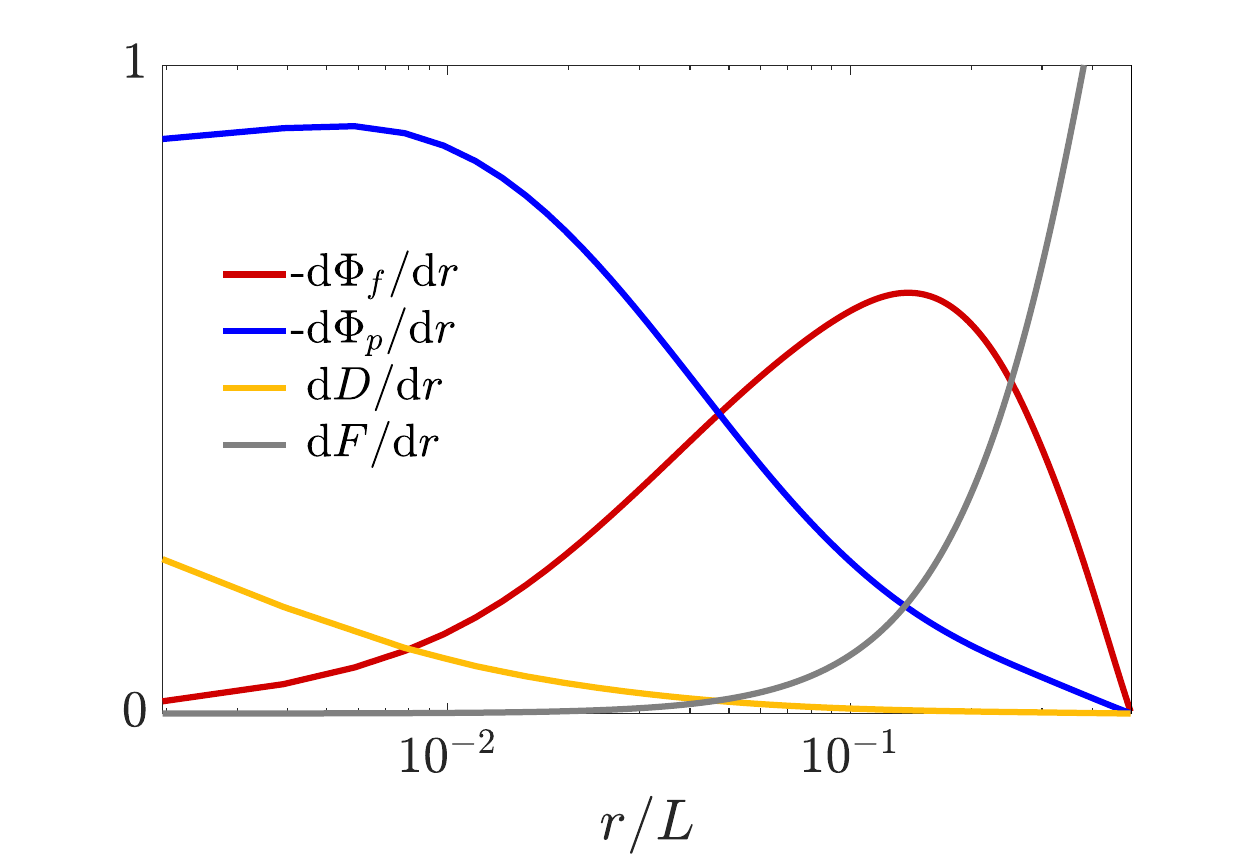}
\includegraphics[width=0.49\columnwidth]{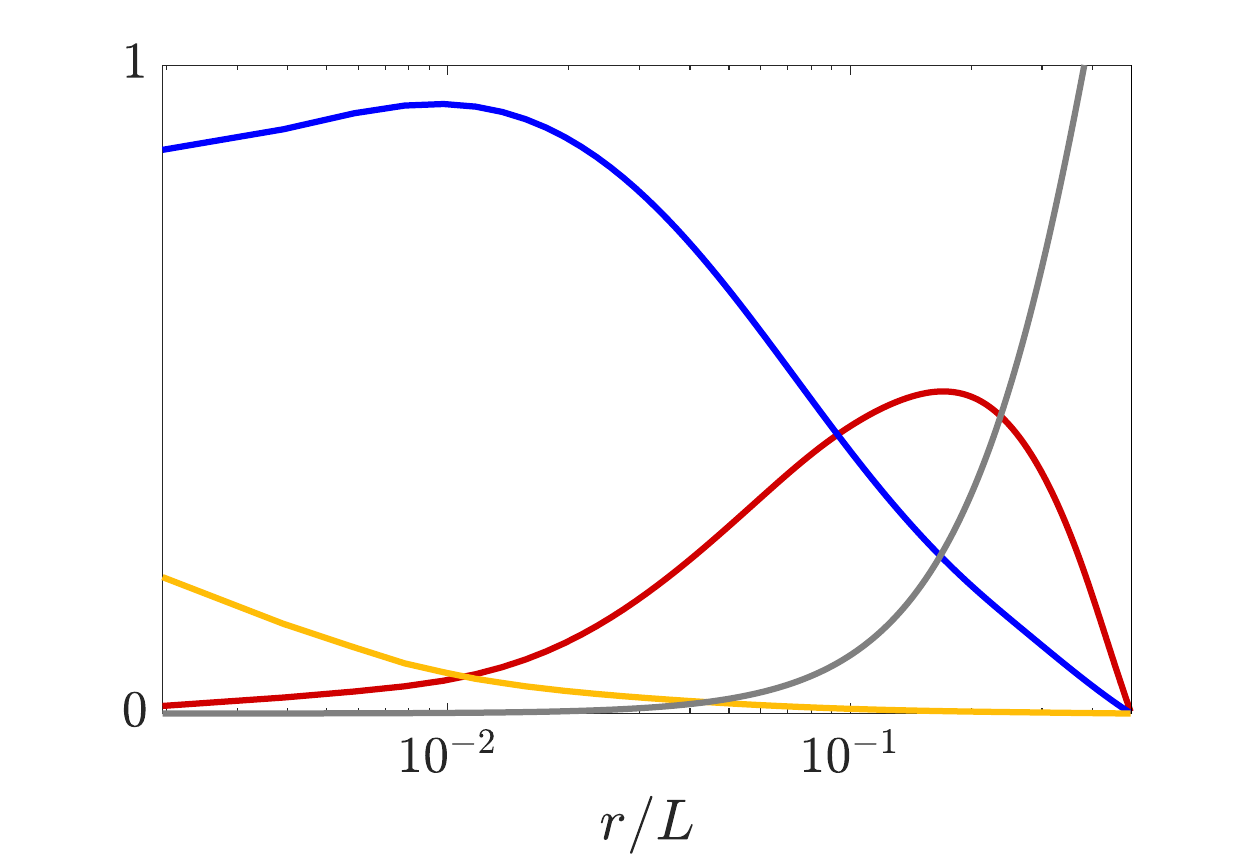}
\includegraphics[width=0.49\columnwidth]{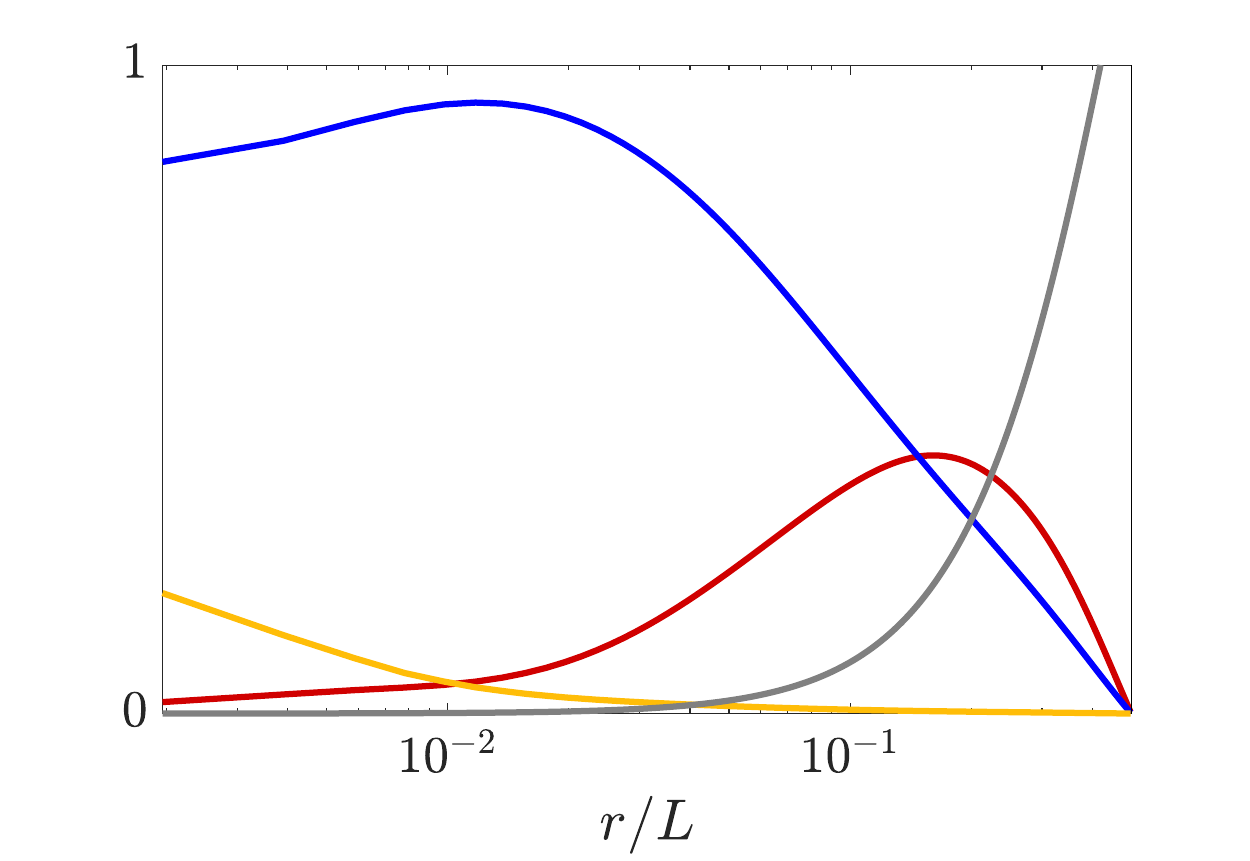}
\caption{Scale-by-scale budget for $\aver{ \delta u_i^2 }$ at $\text{Re}_\lambda \approx 460$. The Deborah number increases from left to right and from top to bottom, being $\text{De}=0$ (top left), $\text{De}=1$ (top right), $\text{De}=3$ (bottom left) and $\text{De}=9$ (bottom right).}
\label{fig:bud_De}
\end{figure}

For completeness, in this section we briefly show the effect of the elasticity of the polymers on the scale-by-scale budget for $\langle \delta u_i^2 \rangle$ (see Eqn. 5 in the main text). This is the real-space equivalent of the Fourier analysis presented by \citet{Marco23} and \citet{singh-rosti-2024}. 
In this section we focus on the largest Reynolds number considered in this work ($\text{Re}_\lambda \approx 460$), but qualitatively the same discussion holds also for the other $Re$.

For the purely Newtonian case ($\text{De}=0$), the top left panel of Fig. \ref{fig:bud_De} statistically characterises the Kolmogorov and Richardson picture of turbulence. At the largest scales energy in injected in the system by the external forcing, and the budget reduces to
\begin{equation}
 \frac{\text{d} F}{\text{d}r} \approx \frac{4}{3} \aver{ \varepsilon_\text{f} }.
\end{equation}
At the intermediate range of scales, where the forcing and viscous terms are negligible, energy is transferred among scales by the nonlinear flux through the classic energy cascade process. Here the energy cascade rate is invariant and equals the dissipation rate, i.e.,
\begin{equation}
  \frac{\text{d} \Phi_\text{f}}{\text{d} r } \approx - \frac{4}{3} \aver{\varepsilon_\text{f}}.
\end{equation}
Eventually, at the smallest scales the viscous contribution $D$ dominates and energy is dissipated by the viscous friction, i.e.,
\begin{equation}
  D  \rightarrow \frac{4}{3} \aver{\varepsilon_\text{f}} \qquad \text{for} \qquad r \rightarrow 0.
\end{equation}

As detailed in the main text, the addition of polymers introduces an alternative route for energy transfer among scales ($\Phi_\text{p}$), and an additional dissipation ($\aver{\varepsilon_\text{p}}$). Fig. \ref{fig:bud_De} shows that the polymers modify the energy cascade in a non trivial way, and that the relevance of the non Newtonian contributions increases with $De$. At the large scales, the energy injected in the system by the external forcing equals the total dissipation:
\begin{equation}
  \frac{\text{d}F}{\text{d}r} \approx \frac{4}{3}  \underbrace{ \left( \aver{ \varepsilon_\text{f}} + \aver{\varepsilon_\text{p} } \right) }_{\aver{\varepsilon_\text{t}}} .
\end{equation}
In the intermediate range of scales where the external forcing and the viscous contribution are negligible, energy is transferred among scales by means of both the nonlinear and polymeric fluxes, and the budget reduces to
\begin{equation}
  \frac{\text{d} \Phi_\text{f}}{\text{d} r} + \frac{\text{d} \Phi_\text{p}}{\text{d} r} \approx - \frac{4}{3} \left( \aver{ \varepsilon_\text{f} } + \aver{ \varepsilon_\text{p} } \right)	.
\end{equation}
In this range, at the largest scales the nonlinear flux takes over, while at the smallest ones the polymeric flux dominates. The relative importance of the two fluxes and the width of the two ranges changes with $\text{De}$. Eventually, at the smallest scales, energy is dissipated away by both the the fluid and the polymers, and we obtain
\begin{equation}
D \rightarrow \frac{4}{3} \aver{\varepsilon_\text{f}} \qquad \text{and} \qquad \frac{\text{d} \Phi_\text{p}}{\text{d} r} \rightarrow \frac{4}{3} \aver{\varepsilon_\text{p}} \qquad \text{for} \qquad r \rightarrow 0. 
\end{equation}

\subsection{Estimation of the $\inv$ invariant}

\begin{figure}
\centering
\includegraphics[width=0.95\columnwidth]{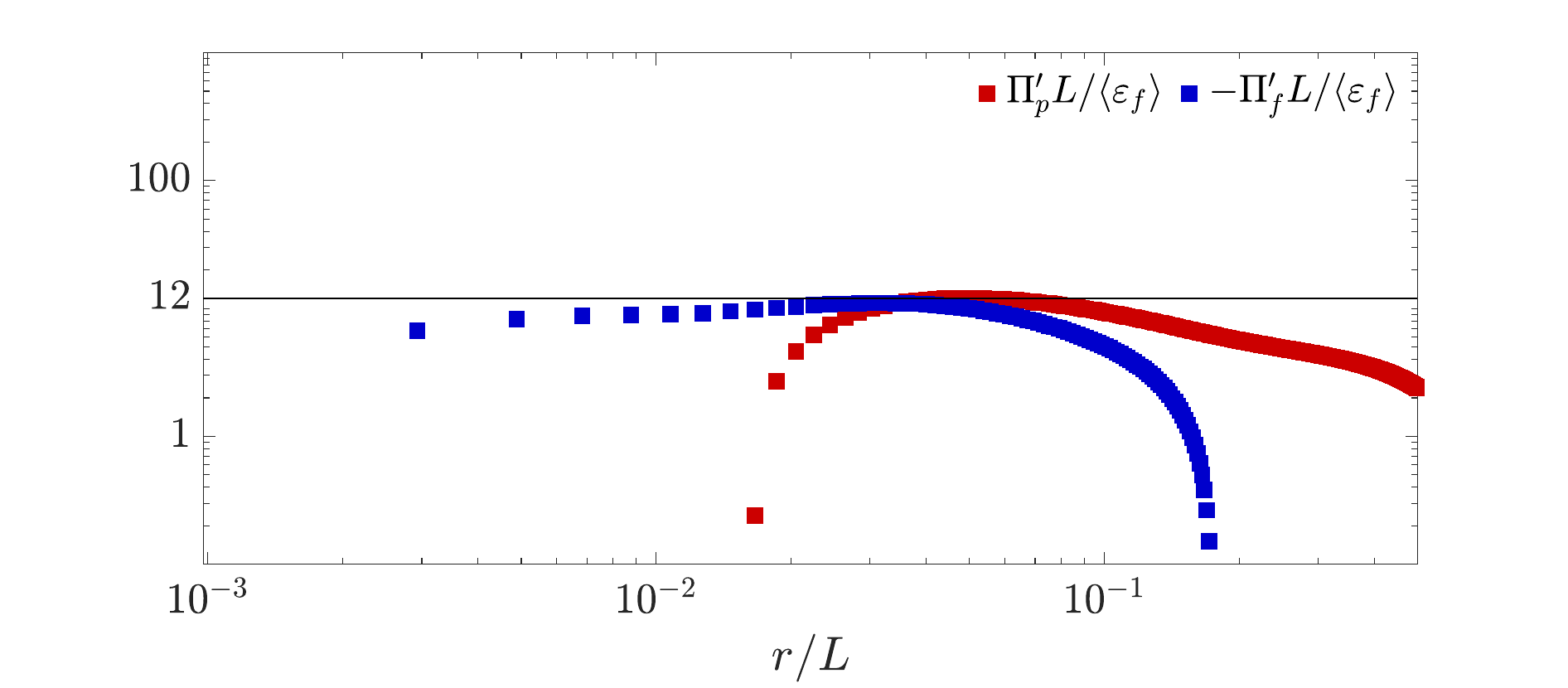}
\includegraphics[width=0.95\columnwidth]{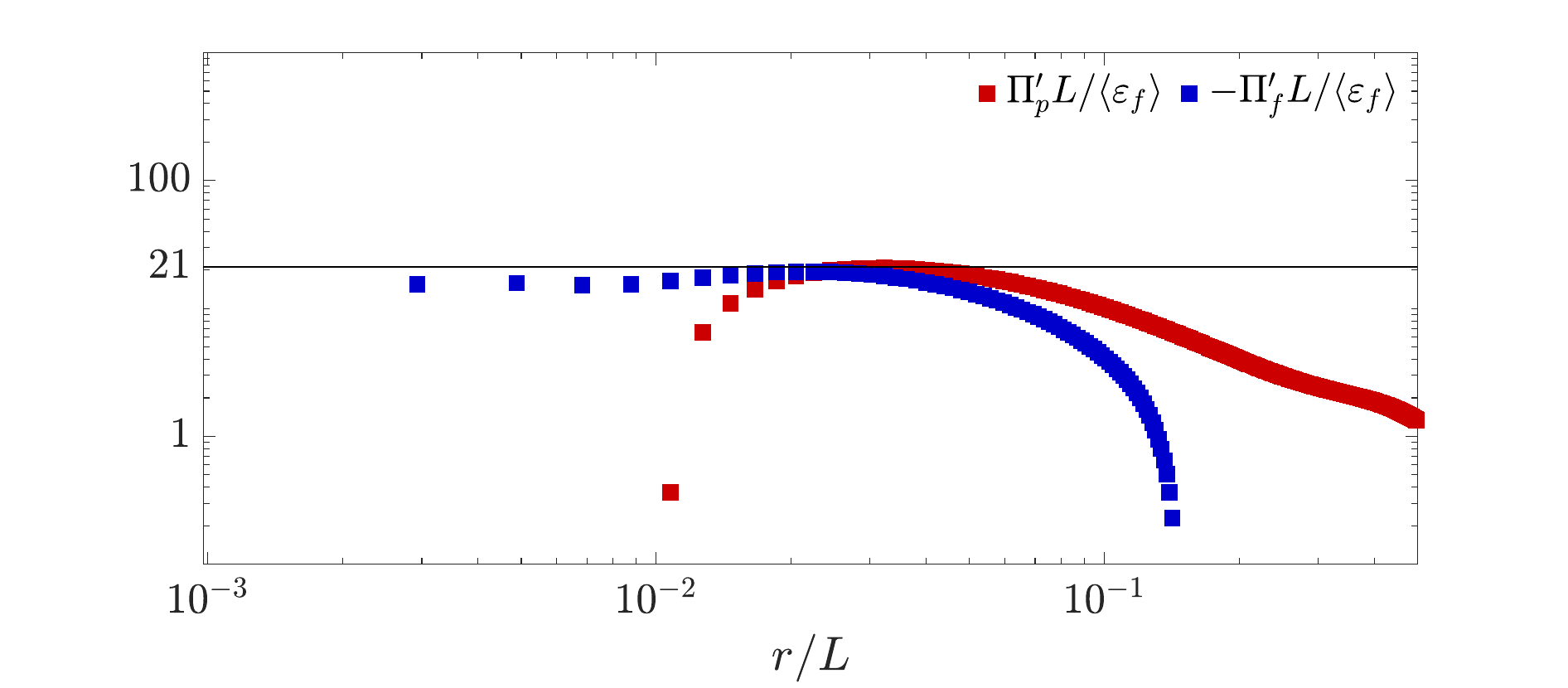}
\includegraphics[width=0.95\columnwidth]{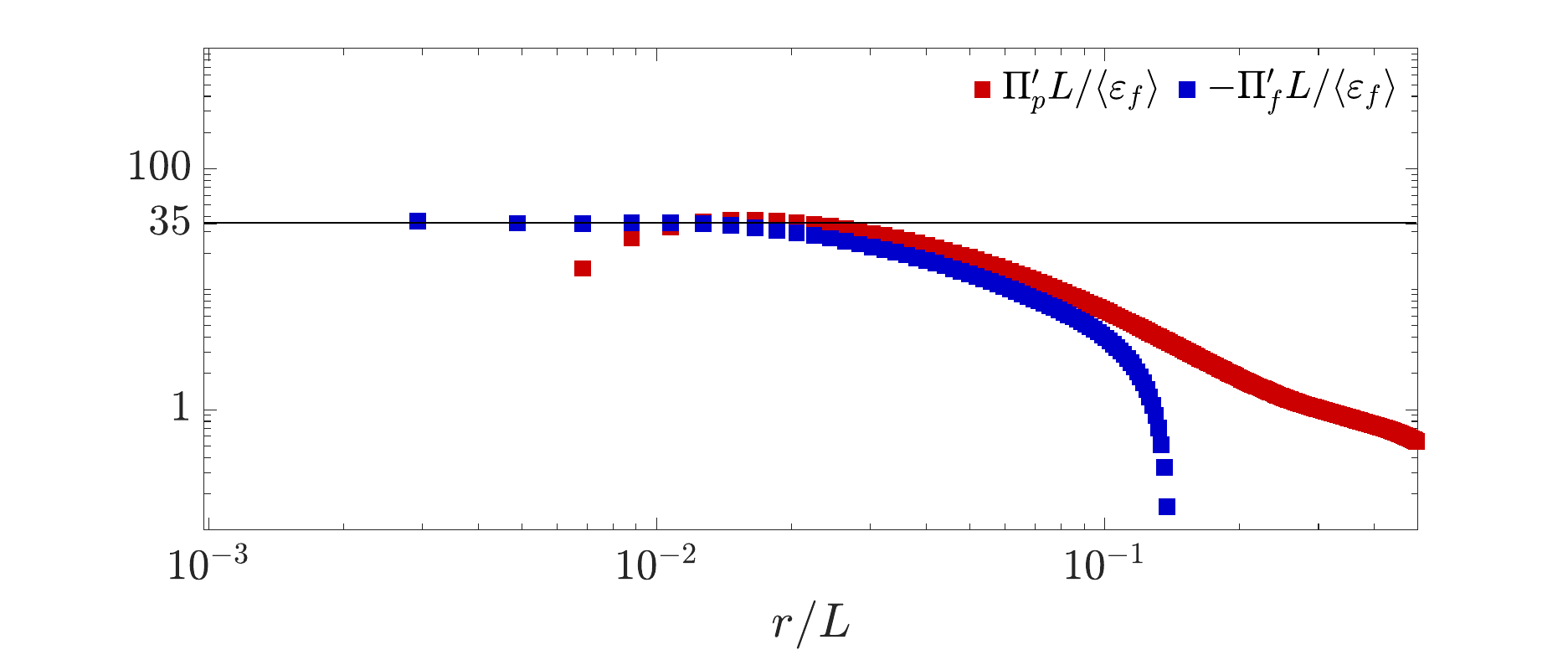}
\caption{Evaluation of the $\inv$ invariant for $\text{De}=1$ and $\text{Re}_\lambda=240$ (top), $\text{Re}_\lambda=320$ (centre), and $\text{Re}_\lambda=460$ (bottom). The red squares refer to $\Pi_\text{p}'$, while the blue squares to $-\Pi_\text{f}'$. The black line tangent to $-\Pi_\text{f}'$ is used to guide the eye to how $\inv$ changes with $\text{Re}_\lambda$.}
\label{fig:d2Phidr2}
\end{figure}

In Fig. 3 of the main text we have shown that the compensated spectrum $\inv^{-2/3} k^{7/3} E(k)$ collapses to approximately the same constant value for different values of $\text{De}$ and $\text{Re}$. As mentioned in the main text, for each case we have estimated the value of $\inv$ directly from the data, exploiting its definition:
\begin{equation}
 \inv = - \frac{\text{d} \Pi_\text{f}}{\text{d} r} =
            \frac{\text{d} \Pi_\text{p}}{\text{d} r}.
\end{equation}
We show in Fig.~\ref{fig:d2Phidr2} that there is indeed a range of scale where $-\Pi_\text{f}'$ and $\Pi_\text{p}'$ exhibit a plateau to (approximately) the same value. For the sake of brevity, here we are considering only $\text{De}=1$ with the Reynolds number varying in the $240 \le \text{Re}_\lambda \le 460$ range. We observe that, the plateau moves towards smaller scales as $\text{R}e$ increases, in agreement with the widening of the intermediate scaling range and the shrinking of the dissipative range of scales (see Fig. 3 of the main text). 

\subsection{The relation between $\inv$ and $\aver{\varepsilon_\text{p}}$}
\label{sec:disspgamma}
\begin{figure}
\centering
\includegraphics[width=0.49\columnwidth]{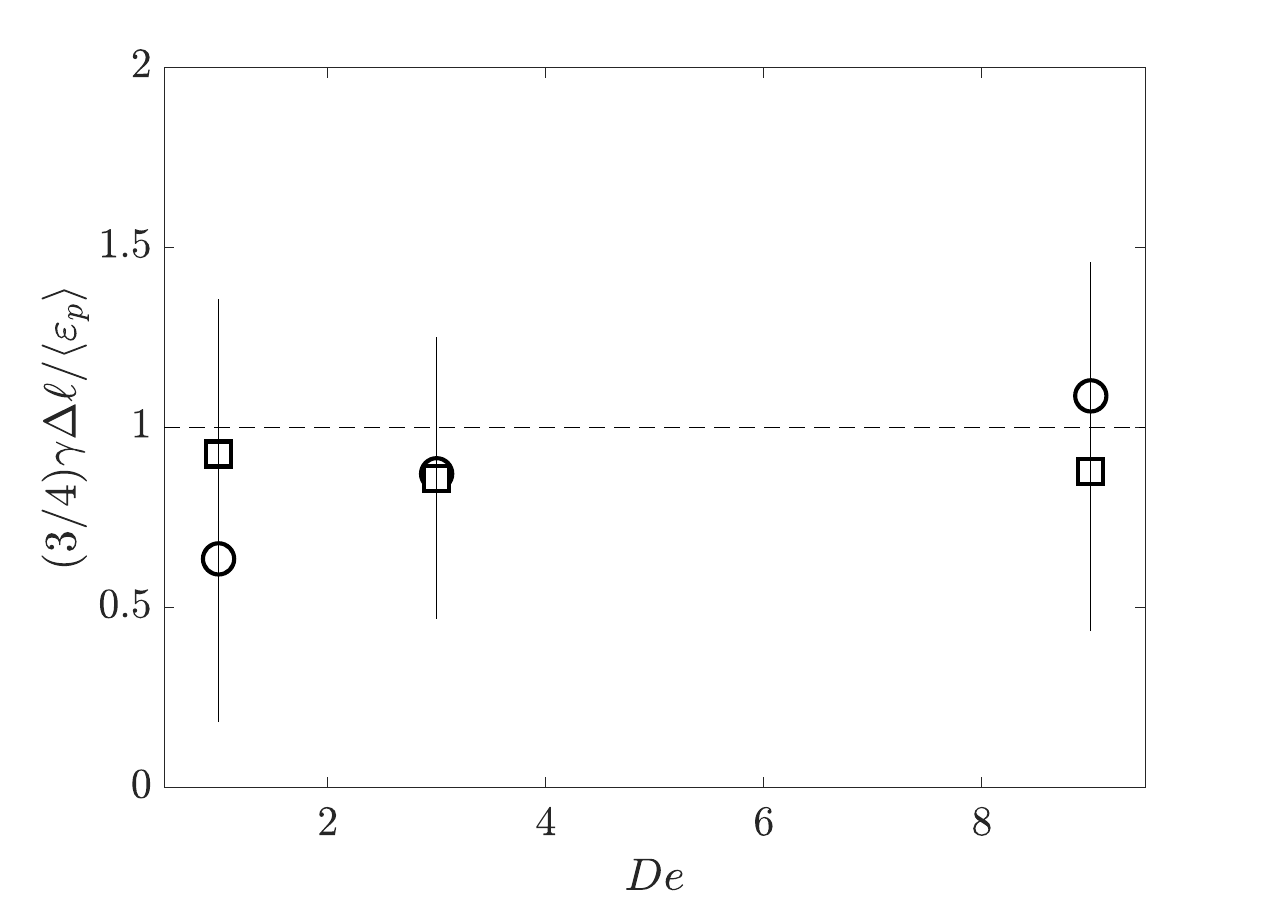}
\includegraphics[width=0.49\columnwidth]{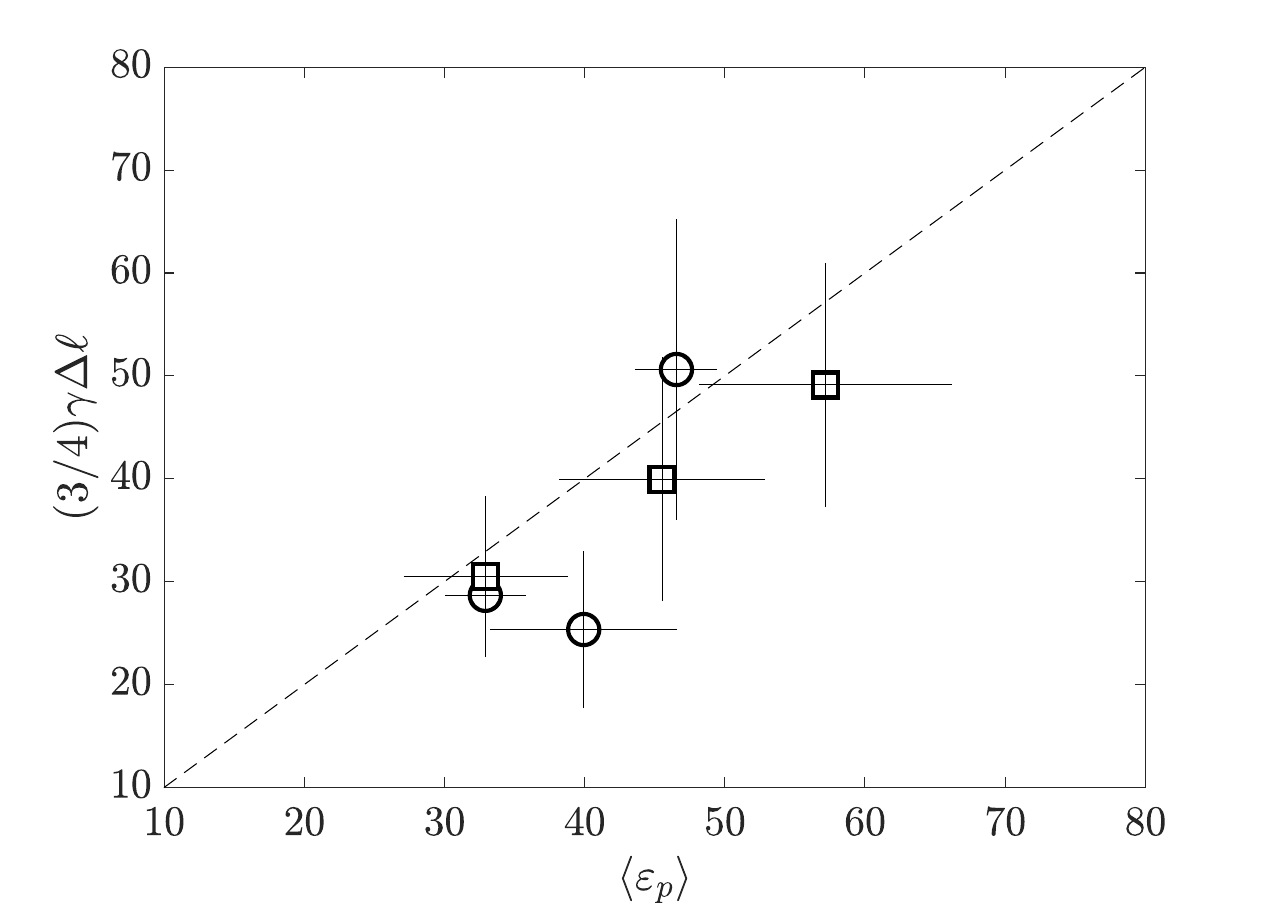}
\caption{The $\inv-\aver{\varepsilon_\text{p}}$ relation for $\text{Re}_\lambda =460$ (circles) and $\text{Re}_\lambda = 240$ (squares). The bars show the standard deviation computed from $6$ different snapshots. The dashed lines represent the predicted behaviour $(3/4)\inv \Delta \ell =\aver{\varepsilon_\text{p}}$.}
\label{fig:gamma_epsp}
\end{figure}

In the main text, we have shown that the invariant $\inv$ is related with the average polymeric dissipation rate $\bra{\dissp}$ by means of a measure of the width of the scaling range $\Delta \ell$. Indeed, in the idealised case with
\begin{equation}
\begin{cases}
  \Pi_\text{p}'  =  \inv \ & \text{for} \ r \in [0, r_b] \\
  \Pi_\text{p}' = 0 \ & \text{for} \ r \in (r_b,L],
\end{cases}
\label{eq:simplified}
\end{equation}
the relation
\begin{equation*}
 \inv = \frac{4}{3} \frac{\aver{\varepsilon_\text{p}} }{\Delta \ell} = \frac{4}{3} \frac{\aver{\varepsilon_\text{p}}}{ r_b }
\end{equation*}
holds.

Although this relation is based on the simplified assumption that the scaling range with $S_2(r) \sim r^{4/3}$ and $E(k) \sim k^{-7/3}$ extends down to the smallest scales, Fig. \ref{fig:gamma_epsp} shows that this prediction is fairly well respected by our numerical data for $\text{Re}_\lambda =240, 460$ and $1 \le \text{De} \le 9$. Here we estimate $\Delta \ell$ as the range of scales where elasticity dominates over inertia, and compute $\inv$ using the top of Eqn.~\eqref{eq:simplified}. To be more accurate, we have approximated $\Delta \ell$ by looking at the range of $r$ where the polymeric flux depleted of the estimated viscous contribution (see \cite{Marco23}), i.e., $\Phi_\text{p}(r) - \aver{\varepsilon_\text{p}}/\aver{\varepsilon_\text{f}} D(r)$, is larger compared to the other terms in the K\'{a}rm\'{a}n--Howarth-Monin--Hill equation (see Eqn. 5 in the main text). The results however only marginally change by looking at the complete $\Phi_\text{p}$. Fig. \ref{fig:gamma_epsp} shows that the ratio $ (3/4) \inv \Delta \ell /\aver{\varepsilon_\text{p}}$ has indeed a value close to one for all the considered cases.

\end{document}